\documentclass[aps,prb,twocolumn,superscriptaddress,showpacs,preprintnumbers,amssymb,graphicx,floatfix]{revtex4-1}
\usepackage{graphicx}
\usepackage{dcolumn}
\usepackage{bm}
\usepackage{natbib}
\usepackage{color}
\usepackage{sidecap}
\usepackage{ulem}
\usepackage{amssymb}
\usepackage{epsfig}
\usepackage{subfigure}

\begin{document}

\title{Magnetization Properties and Vortex Phase Diagram in Cu$_x$TiSe$_2$ Single Crystals}

\date{\today}
\author{P. Husan\'{i}kov\'{a}}
\affiliation{Department of Physics, Drexel University, 3141
Chestnut Street, Philadelphia, Pennsylvania 19104, USA}
\affiliation{Institute of Electrical Engineering, Slovak Academy
of Sciences, D\'{u}bravsk\'{a} cesta 9, SK-841 04 Bratislava,
Slovakia}
\author{J. Fedor}
\affiliation{Institute of Electrical Engineering, Slovak Academy
of Sciences, D\'{u}bravsk\'{a} cesta 9, SK-841 04 Bratislava,
Slovakia}

\author{J. D\'{e}rer}
\affiliation{Institute of Electrical Engineering, Slovak Academy
of Sciences, D\'{u}bravsk\'{a} cesta 9, SK-841 04 Bratislava,
Slovakia}

\author{J. \v{S}olt\'{y}s}
\affiliation{Institute of Electrical Engineering, Slovak Academy
of Sciences, D\'{u}bravsk\'{a} cesta 9, SK-841 04 Bratislava,
Slovakia}

\author{V. Cambel}
\affiliation{Institute of Electrical Engineering, Slovak Academy
of Sciences, D\'{u}bravsk\'{a} cesta 9, SK-841 04 Bratislava,
Slovakia}

\author{M. Iavarone}
\affiliation{Department of Physics, Temple University,
Philadelphia, PA 19122}

\author{S.\ J.\ May} \affiliation{Department Materials Science and
Engineering, Drexel University, 3141 Chestnut Street,
Philadelphia, PA 19104, USA}

\author{G.\ Karapetrov}\email{e-mail: goran@drexel.edu}
\affiliation{Department of Physics, Drexel University, 3141
Chestnut Street, Philadelphia, Pennsylvania 19104, USA}

\begin{abstract}
We have investigated the magnetization properties and flux
dynamics of superconducting Cu$_x$TiSe$_2$ single crystals within
wide range of copper concentrations.  We find that the
superconducting anisotropy is low and independent on copper
concentration ($\gamma\sim1.7$), except in the case of strongly
underdoped samples ($x\leq0.06$) that show a gradual increase in
anisotropy to $\gamma\sim1.9$. The vortex phase diagram in this
material is characterized by broad region of vortex liquid phase
that is unusual for such low-$T_c$ superconductor with low
anisotropy. Below the irreversibility line the vortex solid state
supports relatively low critical current densities as compared to
the depairing current limit ($J_c/J_0\sim10^{-7}$). All this
points out that local fluctuations in copper concentration have
little effect on bulk pinning properties in this system.

\end{abstract}
 \maketitle

$1T-{\mathrm{TiSe}}_{2}$, a quasi-2D layered material with a
trigonal symmetry, has been studied for over 30 years due to the
unconventional nature of its charge density wave (CDW)
state.~\cite{wilson_tise,disalvo_tise,rossnagel_tise,kidd_arpes_tise,monney_tise,vanWezel_prb2011,rossnagel_review}
Recently, the superconductivity was discovered in this system
below 4.15~K by intercalating copper between the van der Waals -
coupled Se-Ti-Se trilayers.~\cite{cava_nphys} Subsequently,
superconductivity was also induced by palladium
intercalation~\cite{morosan_Pd}, as well as by hydrostatic
pressure in pristine
$1T-{\mathrm{TiSe}}_{2}$~\cite{kusmartseva_prl}. Despite the low
superconducting critical temperature $T_c$, the material has
attracted significant attention due to the peculiar nature of the
emergent superconductivity from a semimetallic state above $T_c$,
as well as the coexistence of the superconductivity with the
chiral CDW
state.~\cite{iavarone_chiral,welp_chiral,ishioka_chiral} The
initial studies probing the superconducting phase in
Cu$_x$TiSe$_2$ have yielded diverging results ranging from
multiple superconducting energy gaps~\cite{zaberchik}, to weakly
coupled superconductivity and the presence of spin
fluctuations.~\cite{hillier_uSR} ARPES measurements near the
superconducting transition have shown {\it d}~-~like character of
the emergent density of states near the {\it L} point of the
Brillouin zone at Cu concentrations $x>0.04$ with competing nature
of CDW and superconducting order
parameters.~\cite{hasan_tise_arpes} On the other hand, detailed
specific heat measurements of superconducting Cu$_x$TiSe$_2$ have
shown that the system behaves as a conventional s-wave
superconductor with
$2\Delta/k_BT_c\sim~3.7$.~\cite{li_prl_specheat,kacmarcik_tise} It
is remarkable that for such a simple compound as TiSe$_2$, there
are diverging explanations about the origin of the emergent
superconductivity. The phase diagram of the Cu-doped TiSe$_2$ is
similar to the one of high-temperature superconductors, pnictides,
and heavy fermions, in that the superconducting phase at the
specific doping interval coexists with other correlated electron
states (charge or spin ordering). Considering the relatively
simple lattice structure of the parent compound
$1T-{\mathrm{TiSe}}_{2}$, the system should be suitable for
detailed studies of emergent superconductivity and the evolution
of competing order parameters, CDW and superconductivity.

Besides the intriguing electronic properties very little is known
about the Abrikosov vortex configurations in this superconductor.
The static and dynamic behavior of the vortex lattice, anisotropy
of the superconducting order parameters and vortex lattice phase
diagram could provide further insight into the superconducting
state in Cu$_x$TiSe$_2$ and the specific role that copper plays in
facilitating superconductivity. In this work we study anisotropic
superconducting properties of Cu$_x$TiSe$_2$ single crystals via
bulk magnetization measurements for a range of copper
concentrations spanning from the highly underdoped regime to the
overdoped one. We establish fundamental superconducting parameters
of this system such as upper critical field, coherence length, and
superconducting anisotropy. We establish the vortex phase diagram
of Cu$_x$TiSe$_2$ and we find that the reversibility region is
unexpectedly broad for an extended range of Cu doping values.
Despite the atomic disorder created by intercalated copper atoms,
we observe very low bulk pinning in this material even far from
$T_c$. This points to a uniform amplitude of the superconducting
order parameter across the sample and vortex liquid-like behavior
of the vortex lattice.

Cu$_x$TiSe$_2$ single crystals were grown by means of iodine vapor
transport method in evacuated silica ampoules in a gradient
furnace with the lower temperature part set to 720$^\circ$C and
the temperature gradient of 80$^\circ$°C/m. An average crystal
size was on the order of few mm$^2$ with thickness of several tens
of micrometers. Energy dispersive X-ray spectroscopy (EDS) was
used to establish the quantitative elemental content of the
crystals. The selected single crystals were analyzed for spatial
uniformity of copper concentration by performing EDS analysis at
several points across the sample. The crystals with good spatial
uniformity of copper and sharp superconducting transition were
selected for detailed studies of magnetization properties. The
magnetization measurements were conducted within few weeks from
the single crystal growth as copper tends to migrate with time.
Magnetization measurements were performed in a vibrating sample
magnetometer of PPMS, as well as SQUID magnetometer (Quantum
Design PPMS w/VSM option and MPMS, respectively).

\begin{figure}
\includegraphics[width=3.0in]{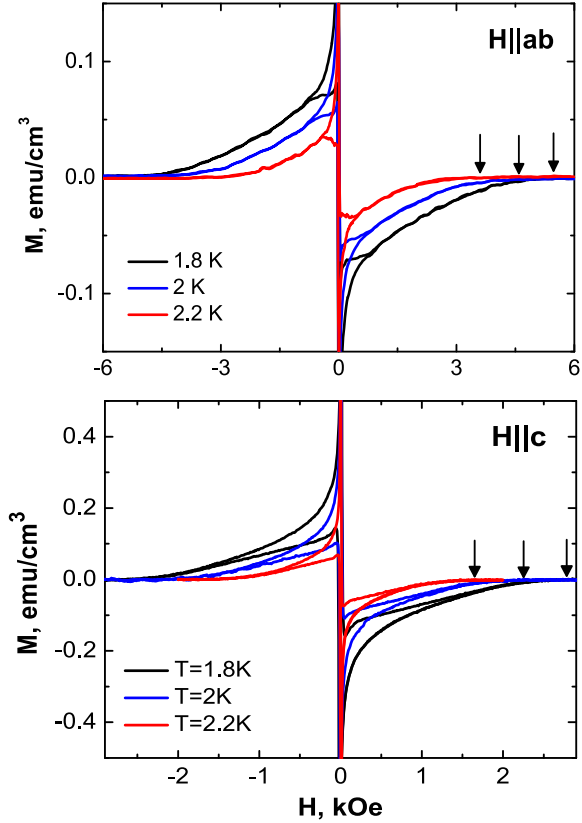}
\caption{(Color online) Magnetization hysteresis loops for
Cu$_{0.058}$TiSe$_2$ sample for $H\parallel ab$ (top) and
$H\parallel c$ (bottom) at different temperatures. The arrows
indicate the transitions from superconducting to normal state,
defining $H_{c2}$ points.}\label{Figure1}
\end{figure}
We studied Cu$_x$TiSe$_2$ single crystals with Cu concentrations
of $x=$~0.058, 0.062, 0.067, 0.085, and 0.090 and these cover the
range from electron underdoped ($x<0.08$) to overdoped ($x>0.08$)
regimes. At normal pressure the superconductivity in
Cu$_x$TiSe$_2$ emerges at Cu doping of around $x=0.04$, reaching
the maximum critical temperature of 4.15K at x=0.08. For copper
concentrations beyond $x=0.08$ the T$_c$ starts to gradually
decrease~\cite{cava_nphys}. Fig.~\ref{Figure1} shows the
representative magnetization loops for Cu$_{0.058}$TiSe$_2$ for
magnetic field applied along the crystallographic planes
($H\parallel ab$) and perpendicular to the planes ($H\parallel c$)
 from which the upper critical fields were
extracted. The hysteresis curves for all samples are characterized
by large reversible regions of the magnetization at higher applied
fields that can be assigned to the magnetization of the Abrikosov
vortex lattice.~\cite{senoussi_review} The irreversibility in
magnetization curves observed at lower fields represents the
effect of vortex pinning.
\begin{figure}
\includegraphics[width=3.0in]{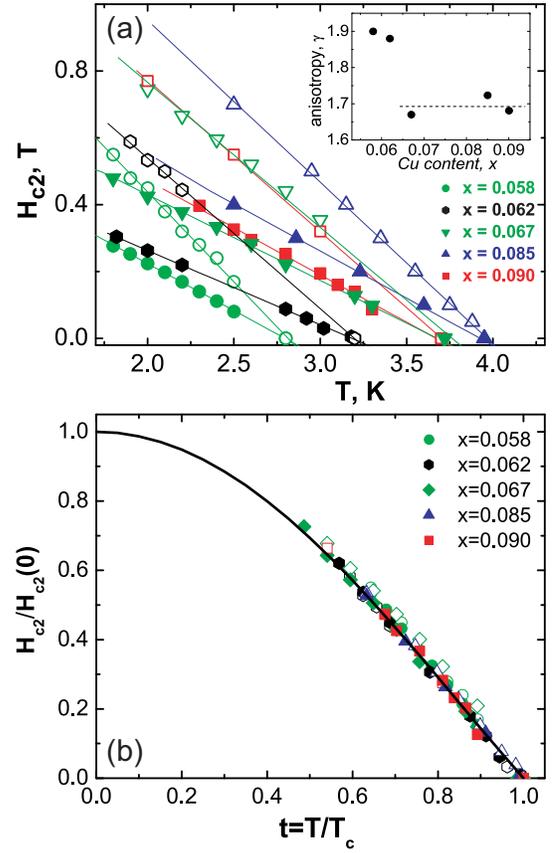}
\caption{(Color online) (a) Temperature dependence of the upper
critical field $H_{c2}$ determined from magnetization measurements
for Cu$_x$TiSe$_2$ samples with five different copper
concentrations. Open symbols are for $H\parallel c$ and filled
symbols are for $H\parallel ab$. The inset shows the values of the
superconducting anisotropy $\gamma=H_{c2}^{ab}(0)/H_{c2}^{c}(0)$
for different copper concentrations {\it x}. (b) Normalized upper
critical field dependence on the reduced temperature for all the
samples in (a) with both directions of the applied field. The full
line is a fit to equation (\ref{whh_fit}).}\label{Figure2}
\end{figure}
From the magnetization curves we extracted the temperature
dependence of the upper critical field $H_{c2}$ along the two
primary axes, perpendicular and parallel to the crystal planes, as
shown in Fig.~\ref{Figure2}a. The linear dependence of the
$H_{c2}(T)$ near $T_c$ is evident for all copper concentrations.
The values of $H_{c2}(0)$ were extracted from the data in
Fig.~\ref{Figure2}a using Werthamer-Helfand-Hohenberg (WHH)
formula~\cite{helfand,*helfand2}
\begin{eqnarray}
H_{c2}(0)=-0.693\cdot T_c\left(\frac{dH_{c2}}{dT}\right)_{T_{c}}
\label{whh}
\end{eqnarray}
This step assumes that the Cu$_x$TiSe$_2$ is an s-wave BCS
superconductor, which is in accordance to recent
results.~\cite{kacmarcik_tise} The data from all samples follow
the empirical formula:
\begin{eqnarray}
\frac{H_{c2}(t)}{H_{c2}(0)}=\left(1-t^2\right)\left(1-at^2\right)
\label{whh_fit}
\end{eqnarray}
with parameter $a=0.3$ that corresponds to approximate WHH
behavior of equation~(\ref{whh}) (i.e.
$H_{c2}(0)=0.7T_cH^{\prime}_{c2}(T_c)$), as shown in
Fig.~\ref{Figure2}b. The perfect scaling of the data for all
copper concentrations signifies that the pair-breaking mechanism
of the magnetic field does not depend on the doping level, i.e. it
does not depend on the amplitude of the underlying CDW state.

\begin{figure}
\includegraphics[width=3.4in]{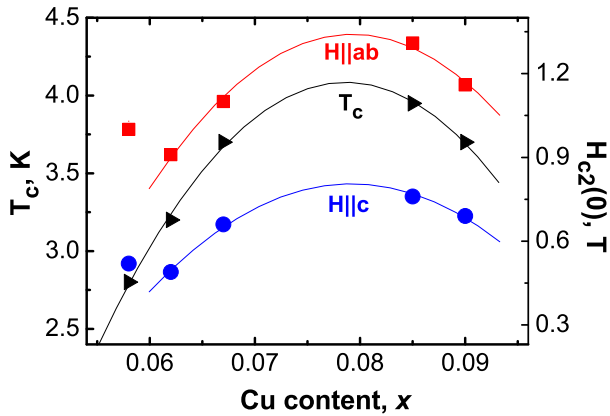}
\caption{(Color online) Dependence of the upper critical fields
and superconducting transition temperature on the Cu concentration
{\it x} in Cu$_x$TiSe$_2$. }\label{Figure3}
\end{figure}

As expected, the $H_{c2}(0)$ varies with Cu concentration {\it x},
closely following the $T_c(x)$ profile (Fig.~\ref{Figure3}). At
close-to-optimal doping level ($x=0.085$) we reach the maximum
$H^c_{c2}(0)$=0.76 T and $H^{ab}_{c2}(0)$=1.31 T. Using the
anisotropic Ginzburg-Landau formulas~\cite{clem_aniGL}:
\begin{eqnarray*}
H^c_{c2}(0)=\frac{\Phi_0}{2\pi\xi^2_{ab}(0)}
\hspace{0.1in};\hspace{0.1in}
H^{ab}_{c2}(0)=\frac{\Phi_0}{2\pi\xi_{ab}(0)\xi_{c}(0)}
\end{eqnarray*}
the superconducting coherence length along the c-axis $\xi_{c}(0)$
and in the ab-plane $\xi_{ab}(0)$ were obtained (here $\Phi_0=
2.07\times10^7$~G$\cdot$cm$^2$~is the flux quantum). Optimally
doped sample with $x=$0.085 shows $\xi_{ab}(0)=$~20.5 nm and
$\xi_{ab}(0)=$~11.9 nm. The anisotropy of the upper critical
field, $\gamma=H_{c2}^{ab}(0)/H_{c2}^{c}(0)$, was found to be 1.7
for $x\geq$~0.067 and independent on the copper concentration.
However, in the highly underdoped regime we found an increase of
the anisotropy to $\gamma\sim1.9$ (inset of Fig.~\ref{Figure2}).
It is interesting to note that compared to other superconducting
transition metal dichalcogenides such as
2H-NbS$_2$~\cite{klein_nbs2},
2H-NbSe$_2$~\cite{coleman_aniNbSe2,levy_aniNbSe2},
Na$_x$TaS$_2$~\cite{wen_aniNaTaS}, and
2H-TaSe$_2$~\cite{fukuyama_aniTaSe2}, Cu$_x$TiSe$_2$ has the
lowest anisotropy of the upper critical field coinciding to one
observed in K$_{0.8}$Fe$_2$Se$_2$
superconductor~\cite{vvm_kfese_apl}. The values of H$_{c2}(0)$
obtained in the slightly underdoped case ($x=0.07$) that were
reported earlier in~\cite{morosan_prb} are consistent with our
dome-like dependence in Fig.~\ref{Figure3}.

The physical origin of the deviation in the upper critical field
and anisotropy in highly underdoped samples may be due to the
confluence of higher electronic anisotropy with lack of available
electronic states to form Cooper pairs. The CDW amplitude is
strong in the underdoped regime and CDW order parameter competes
with superconductivity.~\cite{hasan_tise_arpes} For similar Cu
concentrations ($x=0.055$) an unusual behavior of
magnetoresistance has been reported recently.~\cite{wu_tise}
Moreover, Kusmartseva et al.~\cite{kusmartseva_prl} observed a
sizable suppression of the exponent {\it n} in the
temperature-dependent resistance $R(T) = R_0 + AT_n$ around the
critical pressure of $\sim$3 GPa. This deviation was attributed to
quantum fluctuations in the vicinity of the CDW  quantum critical
point. Similarly, Zaberchik et al.~\cite{zaberchik} observed a
deviation from BCS model in a temperature dependence of a
superfluid density for lower copper concentrations in their
muon-spin rotation experiments. It is quite possible that at lower
Cu concentrations the amplitude of the CDW remains strong despite
the presence of the superconducting order parameter, causing an
increase of the superconducting anisotropy and decrease of
available electronic states participating in superconducting
pairing.

Next, we examine in detail the vortex states in Cu$_x$TiSe$_2$
using DC magnetization hysteresis measurements. The bulk
superconducting critical current densities are determined from the
irreversible part of magnetization loops using the Bean
model~\cite{bean_prl,*bean_rmp}. For $H\parallel c$, the
correspondence between $J_c$ and the width of the magnetization
loop is given by the isotropic Bean
model~\cite{bean_prl,*bean_rmp}, whereas for H$\perp$c we apply
the anisotropic Bean model~\cite{gyorgy_aniBean,*vvm_aniBean} that
takes into account the difference in the superconducting critical
current flowing along and perpendicular to the crystal planes. At
fields of several hundred Oersteds we can assume that the surface
and geometrical barriers do not significantly contribute to the
irreversibility in our samples. We use well established
expressions for critical current densities in anisotropic
superconductors with slab
geometry.~\cite{gyorgy_aniBean,*vvm_aniBean} For $H\parallel c$,
in the case of a rectangular shape of crystal with $b >a >c$~({\it
b, a}, and {\it c} are the length, width, and thickness of the
sample) the in-plane critical current density is given
by~\cite{gyorgy_aniBean,*vvm_aniBean} $J_c^{ab}(H)=\frac{20\Delta
M(H)}{a(1-a/3b)}$~, where {\it a, b} are the dimensions in {\it
cm} and $\Delta M$ is a difference between the magnetization for
decreasing and increasing branch of magnetization loop in
emu/cm$^3$. For magnetic field applied along the {\it ab} plane,
there are two contributions to the supercurrent - parallel to the
Ti-Se layers, $J_c^{ab}$, and perpendicular to the Ti-Se layers,
$J_c^{c}$. In the limit $a,b \gg c\cdot \frac{J_c^{ab}}{3J_c^c}$,
the former is given by $J_c^c=\frac{20\Delta M(H)}{c}$.
\begin{figure}
\includegraphics[width=3in]{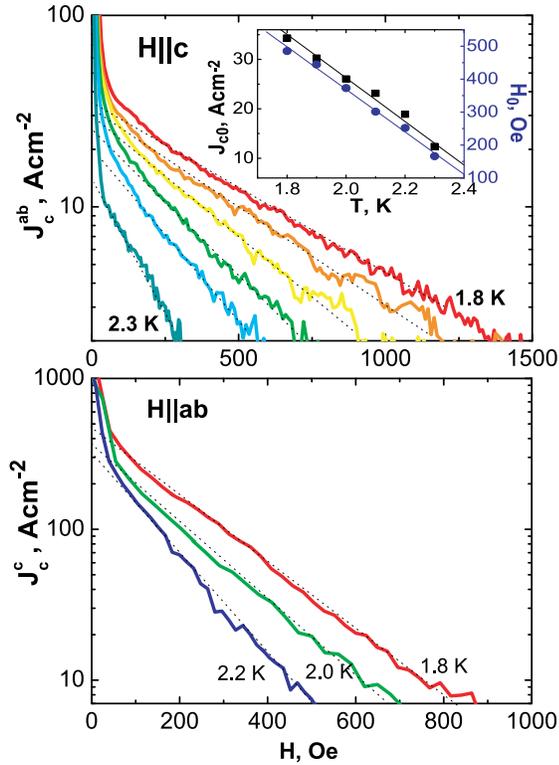}
\caption{(Color online) Magnetic field dependence of the
superconducting critical current density in Cu$_{0.058}$TiSe$_2$
for applied field $H\parallel c$ at $T=$1.8, 1.9, 2.0, 2.1, 2.2,
and 2.3 K (upper panel) and for $H\parallel ab$ at T=1.8, 2.0, and
2.2 K (lower panel). The curves were fit with equation
(\ref{jcvsh}) and temperature dependence of parameters $J_{c0}$
and $H_0$ is shown in the inset.}\label{Figure4}
\end{figure}
An example
of magnetic field dependence of $J_c^c$~and $J_c^{ab}$ for the
sample with x=0.058 are shown in Fig.~\ref{Figure4}. We observe an
exponential dependence that can be expressed using
\begin{eqnarray}
J_c(H,T)=J_{c0}(T)\exp(-H/H_0(T)) \label{jcvsh}
\end{eqnarray}
Extracted values of $J_{c0}(T)$ and $H_0(T)$ show a linear
temperature dependence shown in the inset of Fig.~\ref{Figure4}.
Similar exponential field dependence of the critical current
density was observed in the rest of the samples with different Cu
concentrations. Microscopic examination of the samples by scanning
tunneling microscopy did not reveal defects on the length scale of
the coherence length. The only defects present in the system could
be due to Ti interstitials~\cite{bauer_2ppe} or fluctuation of
copper content on a microscopic scale. Since the average distance
between the Cu atoms is well below the superconducting coherence
length, the intrinsic vortex pinning should be of collective
nature~\cite{larkin_ovchinnikov} caused by the random local
fluctuations of the Cu dopant concentration or atomic Ti
interstitials~\cite{bauer_2ppe}.

\begin{figure}
\includegraphics[width=3in]{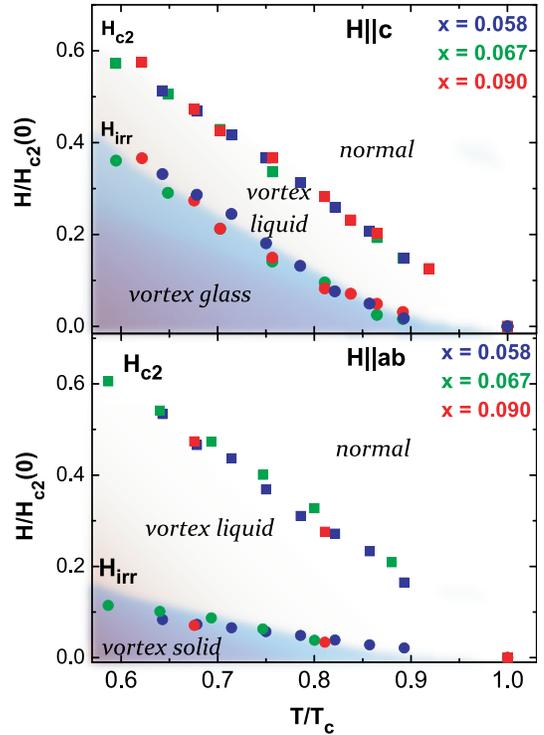}
\caption{(Color online) Vortex phase diagram of Cu$_x$TiSe$_2$ for
magnetic field applied $H\parallel c$ (upper panel) and $\parallel
ab$ (lower panel). }\label{Figure5}
\end{figure}

The relatively large reversible part of the magnetization curve
M(H) spanning up to $H_{c2}$ signifies the presence of a vortex
liquid phase. The boundary between vortex solid and liquid phases,
the irreversibility field $H_{irr}$,~was inferred from the
magnetization hysteresis loops as a field at which upper and lower
magnetization branches merge. As an onset criterium of
reversibility we used the value of $\Delta m \leq 5\times
10^{-7}$~emu. A vortex phase diagram for samples with different
copper concentrations is shown in Fig.~\ref{Figure5}.
The perfect scaling of the $H_{c2}(T)$ and $H_{irr}(T)$ with $T_c$
is further evidence that the underlying nature of the vortex
dynamics in samples with different Cu concentration is the same.
It is remarkable to notice that the reversible region of the
vortex liquid phase is so broad for such a low-$T_c$ material with
relatively low superconducting anisotropy. Normally, the broad
vortex liquid region in the phase diagram is associated with the
existence of thermal fluctuations, a perfect example of which are
the high temperature superconductors. The irrelevance of thermal
fluctuations in our system can be inferred from the estimate of
the Ginzburg number $G\sim3\times10^{-6}$, a value much smaller
than the one found in high-$T_c$ materials ($G\geq 10^{-3}$).\
~\cite{blatter_bible} The estimate of the ratio of the critical
current density to the depairing current density $J_c(H,T)/J_0$,
provides a measure of the magnitude of the quenched disorder
(where depairing current $J_0=\frac{4H_c}{3\sqrt{6}\mu_0\lambda}$,
$\lambda$ is London penetration depth, $H_c$ is thermodynamical
critical field, and $\mu_0$ is permeability of vacuum). In
Cu$_x$TiSe$_2$ this ratio is found to be $\sim 4\times10^{-7}$,
which is much smaller than the one found typically in single
crystal of low-$T_c$ dichalcogenide superconductors such as
NbSe$_2$ ($\sim10^{-3}$). From this we can conclude that the broad
area of the vortex liquid phase in the phase diagram is evidence
of high copper dopant spatial homogeneity that results in
vanishing quenched disorder. It is possible that additional
mechanisms could play a role in reducing the shear modulus of the
vortex lattice, but this issue would need to be addressed through
a combination of microscopic measurements of the local density of
states of the individual vortex line and vortex lattice
configurations~\cite{iavarone_vl_image}.

In conclusion, we have investigated the magnetization properties
and flux dynamics of Cu$_x$TiSe$_2$ single crystals in a wide
range of copper concentrations.  We find that the superconducting
anisotropy is independent on Cu concentration except in the case
of strongly underdoped samples that show a gradual increase in
anisotropy. We establish the vortex phase diagram in this
material. We find a broad, doping-independent region of vortex
liquid phase that is unusual for such low-$T_c$ superconductor
with low anisotropy. Deep in the vortex solid phase the pinning
remains very weak compared to other dichalcogenide
superconductors. This leads us to believe that fluctuations in
copper concentration on a nanometer scale have little effect on
the pinning potential landscape.

\begin{acknowledgments}
This work was supported by Slovak Grant Agency APVV, projects
APVV-0036-11 and VVCE-0058-07, and by the Research \& Development
Operational Program funded by the ERDF, ``CENTE'', ITMS code
26240120011(0.6). Also this work as well as the use of the Center
for Nanoscale Materials and the Electron Microscopy Center at
Argonne National Laboratory were supported by UChicago Argonne,
LLC, Operator of Argonne National Laboratory (``Argonne'').
Argonne, a U.S. Department of Energy Office of Science laboratory,
is operated under Contract No. DE-AC02-06CH11357. Acquisition of
the PPMS was supported by the Army Research Office under DURIP
grant \#W911NF-11-1-0283. M.I. would like to acknowledge the
support of U.S. Department of Energy under Grant No. DE-SC0004556.
\end{acknowledgments}

\bibliographystyle{apsrev}

\end{document}